\documentclass{article}
\usepackage{amssymb}	% Allows use of AMS's mathematical symbols
\usepackage{latexsym}	% Allows use of old latex symbols

{\end{trivlist}}

\begin{document}
\begin{center}

{\Large\bf QUANTUM EFFECTS IN A ROTATING SPACETIME}
\\[1.cm]
Eugen Radu
\\
{\small \emph{Albert-Ludwigs-Universit\"at Freiburg, Fakult\"at f\"ur Physik}, 
\\
\emph{Hermann-Herder-Stra\ss e 3, D-79104 Freiburg, Germany}
\\ email: radu@newton.physik.uni-freiburg.de}
\\[0.2cm]
and 
\\[0.2cm]
Dumitru Astefanesei
\\
{\small \emph{Department of Physics, McGill University, Montreal, QC, H3A 2T8, Canada}
\\
e-mail: astefand@physics.mcgill.ca} 
\\[1cm]
\end{center}

%\date{\today}
%\vspace{1.cm}
\begin{abstract}
The behavior of a arbitrary coupled quantum scalar field 
is studied in the background of the G\"odel spacetime. 
Closed forms are derived for the effective action and 
the vacuum expectation value of quadratic field fluctuations 
by using $\zeta$-function regularization.
Based on these results, we argue that causality 
violation presented in this spacetime can not be removed by
quantum effects.
\end{abstract}
%%%%%%%%%%%%%%%%%%%%%%%%%%%%%%%%%%%%%%%%%%%%%%%%%%%%%%%%%%%%%%%%%%%%%%%%%%%%%%
%%%%%%                      INTRODUCTION
%%%%%%%%%%%%%%%%%%%%%%%%%%%%%%%%%%%%%%%%%%%%%%%%%%%%%%%%%%%%%%%%%%%%%%%%%%%%%%
\section{ Introduction and motivation}

Whether or not the laws of physics permit the existence 
of closed timelike curves (CTCs) is one 
the important problems in the research field of modern theoretical physics. 
A number of familiar spacetimes make it clear that general relativity, 
as it is normally formulated, does not exclude the violation of causality 
in large scale, despite its local Lorentzian character. 

The G\"odel model is not the first but perhaps the best known example of a 
solution of Einstein's field equations in which causality 
may be violated and became a paradigm for causality violations 
in gravitational theory \cite{Godel:1949ga}. 

The most general form for a G\"odel-type homogeneous spacetime was found in 1983 
by Rebou\c{c}as and Tiomno \cite{reboucas}. 
It can be expressed as the direct Riemannian sum $dz^2+d\sigma^2$ of a flat factor and the 
$(2+1)$-dimensional metric
\begin{equation} \label{metric}
d\sigma^2=dr^2+\frac{\sinh^2 m r}{m^2}d\varphi^2-\Big(\frac{4\Omega}{m^2}
\sinh^2(\frac{m r}{2})d\varphi+dt\Big)^2,
\end{equation}
where $\Omega$ and $m$ are constants; 
$m^2$ can be continued to 
negative values, for $m^2 \to 0$ obtaining Som-Raychaudhuri spacetime \cite{Som}. 
The solution originally proposed by G\"odel corresponds to the case $m^2=2\Omega^2$
\begin{equation} \label{metricGodel}
ds^2=dr^2+\frac{\sinh^2\sqrt{2}\Omega r}{2\Omega^2}d\varphi^2+dz^2
-\Big(\frac{2}{\Omega}\sinh^2(\frac{\sqrt{2}\Omega r}{2})d\varphi+dt\Big)^2,
\end{equation} 
where $\Omega \geq 0$ is a constant parameter related to the vorticity 
of the matter, and $-\infty<t,~ z<\infty,~ 0\leq r< \infty, ~0<\varphi\leq 2\pi$.
The source of this geometry is a perfect fluid with constant density 
$\rho$ and no pressure $(p=0)$. Einstein's equation's with a cosmological 
constant $\Lambda$ are satisfied if between $\Omega,~\Lambda$ and 
$\rho$ the following relation holds
\begin{equation}
\Omega^2=-\Lambda=4\pi\rho.
\end {equation}
The study of geodesics showed that this spacetime is geodesically complete 
(and so singularity free) \cite{Chandrasekhar, Novello}. Furthermore, 
because the universe is spacetime homogeneous, 
there are CTCs through every event \cite{Hawking} 
(hence the causal violation is not localized to some small region).

In the last decade many authors have studied features of quantum field 
theory on a spacetime background that contains CTCs but most of 
the papers are dealing with confined causality violating spacetime. 
In these, CTCs are confined within some regions and there exist at least one region
free of them; the regions with CTCs are separated from the well behaved spacetime by
Cauchy horizons.

%there are well behaved initial and final regions and the 
%causality violations are restricted to a region in the middle. 
Surprisingly enough, the study of quantum field theory in G\"odel 
spacetime did not receive enough attention in the literature.
%We can understand this fact  
%Although G\"odel  spacetime does not correspond to the world
%in which we live
Previous works include \cite{Chimento:1989ax}, \cite{Huang:1991eg}, \cite{Radu:1998sk};
 the difficulties in the standard 
formulation of quantum field theory in G\"odel universe have been 
pointed out by Leahy in \cite{Leahy:1982dj} and consist mainly in the absence of a complete 
Cauchy surface and in the incompleteness of the mode solutions to the 
field equations (see also \cite{Novello1993}).

Being a highly symmetric homogeneous spacetime, G\"odel spacetime is 
an excellent model to investigate questions of principle related 
to the quantization of fields propagating on curved background, 
the interaction with a global vorticity and the issues related to 
the lack of global hyperbolicity. Even if one's primary interest 
is in quantum field theory in a confined causality violating spacetime, 
we hope that, by widening the context to G\"odel spacetime, 
one may achieve a deeper appreciation of the theory. 
In particular, one may hopes to attain more general features of a 
quantum field propagating in a nonglobally hyperbolic spacetime, 
whether or not containing a Cauchy horizon. Of particular interest is 
the question whether causality violation which occurs in G\"odel spacetime
implies divergence of vacuum polarization fluctuations and 
may be removed by quantum effects. Thus it seems important to study the 
vacuum polarization for different physical fields in this background.

A different reason to study quantum field theory (QFT)
in this background emerges from the fact
the G\"odel model is the archetypal cosmology exhibiting
the properties associated with the rotation of the universe.
Despite the fact that the cosmological rotation is very small 
by the present observation, it cannot be completely ruled out, 
at least in the early universe \cite{Carneiro:2000yw}
(see also \cite{Obukhov:2000jf} for a up to date discussion 
and a large set of references).

The definition of a quantum field theory on this manifold requires 
some cares; in the following we shall restrict ourselves to 
the Euclidean approach to the quantum field theory, where the 
$\zeta$ function renormalization technique is available. 
The appropriate Euclidean section of G\"odel spacetime is found 
to be the Euclideanized static mixmaster universe. 
We present arguments for the existence of a natural periodicity of 
the euclidean
time, $i.e.$ an intrinsic temperature for a quantum field propagating 
in this background.

The line element (\ref{metric}) has also another interesting property.
We start by considering the most general Taub-NUT-anti de Sitter line-element 
(which can be obtained by analytical continuation of the solutions discussed 
in \cite{Chamblin:1999pz})
written in the form
\begin{equation} \label{nut}
ds^2=\frac{d\chi^2}{V(\chi)}+(\chi^2+n^2)(d\theta^2+f^2(\theta)d\varphi^2)
-V(\chi)\left(dt+4nf^2(\theta/2)d\varphi\right)^2,
\end{equation}
where 
\begin{equation}
V(\chi)=k\left(\frac{\chi^2-n^2}{\chi^2+n^2}\right)+\frac{-2m\chi
+\frac{1}{l^2}(\chi^4+6n^2\chi^2-3n^4)}{\chi^2+n^2}.
\end{equation}
The discrete parameter $k$ takes the values $1, 0$ and $-1$ 
and implies the form of the function $f(\theta)$
\begin{equation}
f(\theta)=\left \{
\begin{array}{ll}
\sin\theta, & {\rm for}\ \ k=1 \\
\theta , & {\rm for}\ \ k=0 \\
\sinh \theta, & {\rm for}\ \ k=-1.
\end{array} \right.
\end{equation}
Here $m$ is the mass parameter, $\chi$ a radial coordinate and $n$ the nut charge.
For vanishing nut charge, the solutions corespond to spherically symmetric ($k=1$) or 
topological ($k=0,-1$) vacuum black holes, with a negative 
cosmological constant $\Lambda=-3/l^2$.
A hypersurface of constant large radius $\chi$ in the four-dimensional Taub-Nut-AdS spacetime
has a metric which is proportional to the three-dimensional line element (\ref{metric})
after the identifications $r=\theta l$; $m=1/l$; $n=\Omega/m^2$.

The Maldacena conjecture \cite{Maldacena:1998re} 
implies that the thermodynamics of a quantum gravity 
with a negative cosmological constant can be modeled by the large $N$ thermodynamics of quantum 
field theory. Therefore the interest in field quantization in the background (\ref{metric}).

The paper is organized as follows: some problems implied by 
the Euclidean approach are discussed in section 2 while in section 3
we compute the local $\zeta$-function. 
The one loop effective action and the vacuum expectation 
value of the field fluctuations are computed in section 4. 
We end drawing some conclusions in section 5.  
%\end{section}

%%%%%%%%%%%%%%%%%%%%%%%%%%%%%%%%%%%%%%%%%%%%%%%%%%%%%%%%%%%%%%%%%%%%%%%%%%%%%%
%%%%%%                    THE EUCLIDEAN SECTION
%%%%%%%%%%%%%%%%%%%%%%%%%%%%%%%%%%%%%%%%%%%%%%%%%%%%%%%%%%%%%%%%%%%%%%%%%%%%%%
\section{The Euclidean section}
Usually there is no well-defined quantum field theory 
in a spacetime containing CTCs.
One way to circumvent some of difficulties of working in spacetime 
which violate causality, 
%emphasized by Hawking in a recent paper 
is to use Hawking's Euclidean quantization procedure \cite{Hawking:1995zi}. 
This approach can be used if some Lorentzian (-causality violating) space has an 
appropriate Euclidean analytic continuation. 
CTCs do not exist in Euclidean space, so one can define a field 
theory on the Euclidean section, and then analytically continue 
to obtain the result valid for the acausal spacetime. 
Note that, for a rotating spacetime, the Wick rotation
is more problematic and generally involves the analytic
continuation of further parameters than the time coordinate.

The use of this approach for G\"odel spacetime
has been originally suggested by Leahy \cite{Leahy:1982dj}.
However, since the G\"odel spacetime is rotating,
 the correct Euclidean formulation of 
a field theory is not immediate and the choice of the 
Euclidean section on which to work is ambiguous.

In Ref.\cite{Radu:1998sk}, the following line element has been proposed 
as describing a rotating gravitational instanton
%%%%%%%%%%%%%%%% line element %%%%%%%%%%%%%%%%%%%%%%%%%%%%%%%%%
\begin{equation}
ds^2=dr^2+\frac{\sin^2\sqrt{2}\Omega r}{2\Omega^2}d\varphi^2+
dz^2+(-\frac{2}{\Omega}\sin^2(\frac{\sqrt{2}\Omega r}{2})d\varphi+dt)^2.
\end{equation}
The source of this geometry is a perfect fluid wit density 
$\rho$, no pressure $(p=0)$ and
\begin{equation}
\Omega^2=\Lambda=4\pi\rho.
\end {equation}
One recover the Lorentzian G\"odel solution by analytically 
continuing $\Omega\to i\Omega$ and $t\to -it$. 
Note the analogy with the case of rotating black 
hole where a real Euclidean metric is usually obtained by 
supplementing the analytic continuation $t\to it$ of 
the time coordinate $t$ by a further transformation 
$J\to iJ$ (where $J$ is the real angular momentum \cite{Gibbons:1977ue}). 
However, in contrast with asymptotic meaning of rotation 
and other quantities of a rotating black hole, 
in this case the rotation has a well defined local 
character being related to the matter contents of the universe.

%Thus we can always perform all the calculation in the
Thus we always perform the calculation in the  
Euclidean geometry; analytically continuing at the end 
of the calculation will yield the results for the acausal Lorentzian spacetime.

In order to clarify the relation between the Lorentzian and Euclidean sections
%how the G\"odel spacetime can be analytically continued 
%to the Euclidean section 
we use a result of Rooman and Spindel 
who clarified the problem of the global embedding of the Lorentzian 
section of the G\"odel spacetime \cite{Rooman:1998xf}.

Let $(Z^1,Z^2,Z^3,Z^4,Z^5,Z^6,Z^7,Z^8)$ be global complex coordinates 
on $C^8$ and let $C^8$ be endowed with flat metric
\begin{equation}
ds^2=(dZ^1)^2+(dZ^2)^2+(dZ^3)^2+(dZ^4)^2+(dZ^5)^2-(dZ^6)^2-(dZ^7)^2-(dZ^8)^2.
\end{equation}
We define the complexified G\"odel spacetime $M^C$ as the algebraic variety 
in $C^8$ determined by the four polynomials
\begin{eqnarray}
-(Z^1)^2-(Z^2)^2+(Z^5)^2+(Z^6)^2 & = & \frac{4}{\Omega ^2},
\nonumber\\
(Z^3)^2-(Z^7)^2-(Z^8)^2 & = & \frac{1}{2\Omega ^2},\\
Z^7 & = & \frac{\Omega}{2\sqrt2}(Z^1Z^5-Z^2Z^6),\nonumber\\
Z^8 & = & \frac{\Omega}{2\sqrt2}(Z^2Z^5+Z^1Z^6),\nonumber
\end{eqnarray}
where $\Omega$ is an complex parameter determined by the matter content. 

The Lorentzian and Euclidean sections of interest, denoted by $M^L$ and 
$M^E$ are the subsets of $M^C$ stabilized by the respective antiholomorphic involutions
%
%%%%%%%%%%%%%%%% antiholomorphic involutions %%%%%%%%%%%%%%%%%%%%%%%%%%%%%%%%
%
\begin{equation}
J_L:(Z^1,Z^2,Z^3,Z^4,Z^5,Z^6,Z^7,Z^8)\to 
(\overline Z^1,\overline Z^2,\overline Z^3,
\overline Z^4,\overline Z^5,\overline Z^6,\overline Z^7,\overline Z^8),
\end{equation}
supplemented by the condition $\Omega \to \overline \Omega$, and
\begin{equation}
J_E:(Z^1,Z^2,Z^3,Z^4,Z^5,Z^6,Z^7,Z^8)\to (\overline Z^1,\overline Z^2,
-\overline Z^3,\overline Z^4,-\overline Z^5,-\overline Z^6,\overline Z^7,\overline Z^8),
\end{equation}
with $\Omega\to -\overline \Omega$, such that $J_L$ leaves 
$M^L\subset M$ invariant, $J_L(M^L)=M^L$ while $J_E$ leaves 
$M^E\subset M$ invariant, $J_E(M^E)=M^E$. Clearly $M^L$ and 
$M^E$ are real algebraic varieties; on $M^L$, $Z^i$ are real 
for all $i$; on $M^E$, $Z^i$ are real for $i=1,2,4,7,8$ while 
$Z^3, Z^5$ and $Z^6$ are purely imaginary. 
The induced metric has a number of different representation
depending on the choice of coordinates. 
An explicit embedding of $M^L$ in terms of the usual 
G\"odel coordinates (which cover the entire variety) reads

%%%%%%%%%%%%%%%% embeddings %%%%%%%%%%%%%%%%%%%%%%%%%%%%%%%%
\begin{eqnarray}
Z^1 & = & \frac{2}{\Omega}\sinh\frac{\sqrt 2 \Omega r}{2}\cos(\varphi-\frac{\Omega t}{2}),
\nonumber\\
Z^2 & = & \frac{2}{\Omega}\sinh\frac{\sqrt 2 \Omega r}{2}\sin(\varphi-\frac{\Omega t}{2}),
\nonumber\\
Z^3 & = & \frac{1}{\sqrt 2\Omega}\cosh\sqrt 2 \Omega r,
\nonumber\\
Z^4 & = & z, \\
Z^5 & = & \frac{2}{\Omega}\cosh\frac{\sqrt 2 \Omega r}{2}\cos(\frac{\Omega t}{2}),
\nonumber\\
Z^6 & = & \frac{2}{\Omega}\cosh\frac{\sqrt 2 \Omega r}{2}\sin(\frac{\Omega t}{2}),
\nonumber\\
Z^7 & = & \frac{2}{\sqrt 2\Omega}\sinh\frac{\sqrt 2 \Omega r}{2}\cos \varphi,
\nonumber\\
Z^8 & = & \frac{2}{\sqrt 2\Omega}\sinh\frac{\sqrt 2 \Omega r}{2}\sin \varphi,
\nonumber
\end{eqnarray}
while the embedding of $M^E$ is
\begin{eqnarray}
Z^1 & = &\frac{2}{\Omega}\sin\frac{\sqrt 2 \Omega r}{2}\cos(\varphi-\frac{\Omega t}{2}),
\nonumber\\
Z^2 & = & \frac{2}{\Omega}\sin\frac{\sqrt 2 \Omega r}{2}\sin(\varphi-\frac{\Omega t}{2}),
\nonumber\\
Z^3 & = &\frac{1}{\sqrt 2\Omega}\cos\sqrt 2 \Omega r,
\nonumber\\
Z^4 & = &z, \\
Z^5 & = & \frac{2}{\Omega}\cos\frac{\sqrt 2 \Omega r}{2}\cos(\frac{\Omega t}{2}),
\nonumber\\
Z^6 & = & \frac{2}{\Omega}\cos\frac{\sqrt 2 \Omega r}{2}\sin(\frac{\Omega t}{2}),
\nonumber\\
Z^7 & = & \frac{2}{\sqrt 2\Omega}\sin\frac{\sqrt 2 \Omega r}{2}\cos \varphi,
\nonumber\\
Z^8 & = & \frac{2}{\sqrt 2\Omega}\sin\frac{\sqrt 2 \Omega r}{2}\sin \varphi,
\nonumber
\end{eqnarray}
where
\begin{eqnarray}
(Z^1)^2+(Z^2)^2+(Z^5)^2+(Z^6)^2 & = & \frac{4}{\Omega ^2},
\nonumber\\
(Z^3)^2+(Z^7)^2+(Z^8)^2 & = & \frac{1}{2\Omega ^2},
\\
Z^7 & = & \frac{\Omega}{2\sqrt2}(Z^1Z^5-Z^2Z^6),
\nonumber\\
Z^8 & = & \frac{\Omega}{2\sqrt2}(Z^2Z^5+Z^1Z^6).
\nonumber
\end{eqnarray}
It follows that $M^L$ does not admit a global foliation with 
spacelike hypersurfaces \cite{Rooman:1998xf}; the isometries of $M^C$ are clear 
from the above construction.
On $M^E$ the globally-defined Killing vector $\partial _t$ 
generates a $U(1)$ isometry group of rotations; thus the 
Euclidean G\"odel time $t$ is periodic with a period $4\pi/\Omega$. 
A periodicity $\beta \ne \frac{4\pi}{\Omega}$ of the Euclidean coordinate 
$t$ implies a Dirac's delta singularity in the curvature of the manifold at $r=0$.

In \cite{Radu:1998sk} the line element (2.1) was obtained considering it as the direct 
sum of the metric $\boldmath{g_1}$ on a 3-dimensional space $S^3$ of 
constant curvature, 
$dl^2=\gamma _{ab} \boldmath{\sigma}^a\boldmath{\sigma}^b$ 
(where the $\boldmath{\sigma}^a$'s are the basis one-forms on 
the three sphere satisfying the structure relations 
$d\boldmath{\sigma}^a=\frac{1}{2}\epsilon^a_{bc}\boldmath{\sigma}^b
\boldmath{\sigma}^c)$ and the metric $\boldmath{g_2}$ 
defined by $ds_2^2=dz^2$ on the 1-dimensional manifold $R$ defined 
by the coordinate $z$. Here $\epsilon ^a_{bc}$, 
components of the totally antisymmetric tensor, 
are the structure constants for the rotation group $SO(3)$ 
and the $\gamma _{ab}=l_a^2\delta_{ab}$ are constant of the space. 
Using the Euler angle-parametrization, the basis forms are
%%%%%%%%%%%%%%%% Euler angle-parametrization %%%%%%%%%%%%%%%%%%%%%%%%%%%%%%%%
\begin{eqnarray}
\boldmath{\sigma}^1 & = & -\sin\Psi d\theta+\cos\Psi \sin\theta d\varphi,
\nonumber\\
\boldmath{\sigma}^2 & = & -\cos\Psi d\theta+sin\Psi \sin\theta d\varphi, \\
\boldmath{\sigma}^1 & = & d\Psi + \cos\theta d\varphi,\nonumber
\end{eqnarray}
with $0<\theta \le \pi,~ 0<\varphi \le 2\pi,~ 0<\Psi \le 4\pi$. 
Taking $l^2_1=l^2_2=l_3^2/2=1/2\Omega ^2$, 
and considering the coordinates transformations $\theta=\sqrt{2}\Omega r, 
t=-\frac{1}{\Omega}(\varphi +\Psi)$ we obtain the line element (3), 
the coordinate range in order to avoid a Dirac's delta singularity 
 being $0<\sqrt{2}\Omega r \le \pi,~0<t\le 4\pi/\Omega, 
~0<\varphi \le 2\pi,~ -\infty<z<\infty$. 

Given the thermodynamical principle that the 
temperature $T$ is inversely related to the period $\beta,~T=\beta ^{-1}$, we might 
therefore expect that the temperature of a quantum field propagating in G\"odel 
spacetime would be $\Omega/4\pi$ refereed to the Killing 
vector generated by the Lorentzian time $t$. 
Thus it corresponds to the heat bath and will
be in a mixed quantum state. 
A free field would propagate straight through the
 heat bath and not notice its existence but an interacting 
field will be affected and will lose quantum coherence to the heat bath. 

For both Hawking and Unruh effects, temperature emerges from information 
loss associated with real and accelerated-observer horizons, respectively. 
The case of the G\"odel spacetime is rather different, given the different global structure. 
This spacetime does not present a causal horizon that hides information; 
however there is an information loss associated with CTCs.
According to a more general argument presented in \cite{Hawking:1995zi}, 
one might expect loss of quantum coherence whenever one has CTCs, 
because there will be a part of the quantum state that one doesn't measure initially or finally.

Also, it is worth remarking that the line element (2.1) corresponds to the Euclidean section of a 
static Taub universe \cite{Taub:1951ez} written in a slightly modified form and some finite 
temperature expressions in the G\"odel case can be in principle read off from 
the zero temperature static Taub results 
(a similar relation exists $e.g.$ between the QFT 
in cosmic string spacetime and Rindler spacetime  \cite{Iellici:1998ce}).
QFT in Taub spacetime has been  discussed in the past 
by many authors (\cite{Critchley:1981am},\cite{Shen:1985ir}, \cite{Stylianopoulos:1989sk}). 

Further interest in this line-element emerges recently from 
the AdS/CFT correspondence,
since the squashed three sphere is the boundary of a euclidean four dimensional
Taub-Nut-AdS spacetime.
Recent papers on field quantization on this geometry are \cite{Dowker:1999pi, DeFrancia:2001xm}. 

%\end{section}

%%%%%%%%%%%%%%%%%%%%%%%%%%%%%%%%%%%%%%%%%%%%%%%%%%%%%%%%%%%%%%%%%%%%%%%%%%%%%%
%%%%%%                   LOCAL ZETA FUNCTION
%%%%%%%%%%%%%%%%%%%%%%%%%%%%%%%%%%%%%%%%%%%%%%%%%%%%%%%%%%%%%%%%%%%%%%%%%%%%%%
\section{Local zeta function}
A very convenient method to compute the Euclidean Green function, 
the vacuum fluctuation and the one-loop renormalized stress tensor 
for a quantum scalar field propagating in the background (\ref{metric}) 
is to use the formalism of ``the direct local 
$\zeta$-function approach'' \cite{Moretti:1997qn}. 
The local $\zeta$-function related to the elliptic operator $A$ positive 
definite on the Euclidean manifold $M^E$, can be defined as the analytical
continuation of the series
\begin{equation}
\zeta (s,x|A)={\sum_N} ' \lambda_N^{-s}\Phi^* _N(x)\Phi _N(x),
\end {equation}
(where ' indicates that possible null eigenvalues are omitted). 
For a scalar field, 
the global $\zeta$-function can be obtained by 
integrating the local $\zeta$-function  $\zeta (s,x|A)$
\begin{equation}
\zeta (s|A)=\int_{M^E}d^4x\sqrt{g}\zeta(s,x|A),
\end {equation}
where, through the spectral representation, $\Phi_N(x)$ is a 
complete series of normalized eigenvectors of the elliptic 
differential second-order selfadjoint operator $A=-\nabla ^\mu\nabla _\mu+M^2+\xi R$ 
with eigenvalues $\lambda_N$. 
Here $\xi$ is a parameter which fixes the coupling of the field to the gravity 
by means of the scalar curvature $R$. Actually, when the manifold is non-compact, 
only the local zeta function has a precise mathematical meaning, since the integration 
requires the introduction of cutoffs or smearing functions to avoid 
divergences \cite{Moretti:1997qn}.

%%%%%%%%%%%%% eigenfunctions %%%%%%%%%%%%%%%%%%%%%%%%%%%%%
The eigenfunctions of the operator $A$ in the Euclideanized G\"odel spacetime are just 
the normalized rotation matrices \cite{Radu:1998sk}
\begin{equation}
\Phi _N=\sqrt{\frac{\Omega^3(2J+1)}{8\pi^3}}
D_{mm'}^J(\sqrt{2}\Omega r)e^{i((m-m')\varphi+k_zz-m\Omega t)},
\end{equation}
$D_{mm'}^J(\sqrt{2}\Omega r)$ being the Wigner functions. 
The corresponding eigenvalues are
\begin{equation}
\lambda_N=2\Omega^2\left(J(J+1)+\frac{k_z^2+M^2+\xi R}{2\Omega^2}-\frac{m^2}{2}\right),
\end{equation}
where $N={J, m, m',k_z}$, $-\infty <k_z<\infty$; $J$ takes all 
values of positive integers and half integers and $m,m'=-J,-J+1,...,J-1,J$ 
(notice $\lambda_N$ has an $m'$ degeneracy). 
To compute $\zeta (s,x|A)$ we first use the sum rule \begin{displaymath}
\sum_{m''} D_{mm''}^J(D_{m'm''}^J)^*=\delta _{mm'}
\end{displaymath} 
to obtain
%%%%%%%%%%%%% local zeta %%%%%%%%%%%%%%%%%%%%%%%%%%%%%
\begin{equation}
\zeta(s,x|A)=\frac{\Omega ^4}{(4\pi)^2} 
\frac{\Gamma(s-\frac{1}{2})}{\sqrt{2\pi}\Gamma (s)} 
(\frac{2}{\Omega^2})^s
\sum_{J=0}^{\infty}
\sum_{m=-J}^{J}\frac{(2J+1)}
{
\Big(J(J+1)+\frac{m^2+\xi R}{2\Omega^2}-\frac{m^2}{2}
\Big)^{s-\frac{1}{2}}}.
\end{equation}
Therefore, due to the symmetry of spacetime, the local zeta function 
do not depend on $x$ and so the corresponding fluctuations of the field. 
Somewhat similar series occur when discussing quantum effects in Taub spacetime; 
however, the approach presented here and the final results in this paper differs 
from those in existing literature on the same subject
\cite{Critchley:1981am}-\cite{ Dowker:1999pi}. 

The summation over $m$ can be performed by the expansion
\begin{equation}
\zeta(s,x|A)=\frac{\Omega ^4}{(4\pi)^2} \frac{1}{\sqrt{2\pi}\Gamma(s)}
(\frac{2}{\Omega^2})^s
\sum_{k=0}^{\infty}
\sum_{J=0}^{\infty}
\sum_{m=-J}^{J}
\frac{(-1)^k(2J+1)m^{2k}}{2^kk!\Big(J(J+1)+\frac{m^2+\xi R}{2\Omega^2}-\frac{m^2}{2}
\Big)^{s+k-\frac{1}{2}}}\frac{\Gamma(-s+\frac{3}{2})}{\Gamma(-s-k+\frac{3}{2})}
,
\end{equation}
and making use of the Plana sum formula which says that, for a function $f(m)$
\begin{equation}
\sum_{m=j}^\infty f(m)=\frac{f(j)}{2}+\int_{j}^{\infty} f(x)dx+i\int_{0}^
{\infty}\frac{dt}{e^{2\pi t}-1}[f(j+it)-f(j-it)].
\end{equation}
We find
\begin{eqnarray}  \label{f1}
\lefteqn{ \zeta(s,x|A)=\frac{\Omega ^4}{(4\pi)^2} 
\frac{\Gamma(s-1/2)}{\sqrt{2\pi}\Gamma(s)}(\frac{2}{\Omega^2})^s
\Bigg(
\sum_{n=1}^{\infty}\frac{n^2}{(n^2+\sigma)^{s-1/2}}+{} } \\
& & {}+\sum_{k=1}^{\infty}\frac{(-1)^k}{2^kk!}\frac{\Gamma
(-s+\frac{3}{2}}{\Gamma(-s-k+\frac{3}{2})}
\Big[\frac{1}{2k+1}I(s+k-\frac{1}{2};2k+1;\sigma)+{}
\nonumber\\
& & {}+I(s+k-\frac{1}{2};2k;\sigma)+\sum_{p=0}^{k-1}  {2k \choose 2p+1}  2^{2k-2p-1}
\frac{B_{2k-2p}}{k-p}I(s+k-\frac{1}{2};2p+1;\sigma)
\Big]
\Bigg)\nonumber
,
\end{eqnarray}
where $B_k$ are Bernoulli numbers and $\sigma=2\Omega ^2(M^2+\zeta R)-1$ 
(for a massless conformally coupled scalar field $\sigma=-\frac{1}{3}$; 
while for $M=0, \xi =0, \sigma=-1$; in this case the value $n=1$ corresponds 
to a null eigenvalues and is omitted). 
In (\ref{f1}) we have defined 
\begin{equation}
I(q,r,\sigma)=\sum_{n=1}^{\infty}\frac{n(n-1)^r}{(n^2+\sigma)^q}. 
\end{equation}
Some properties of the functions $I(q,r,\sigma)$ are discussed in the Appendix; 
they possess poles at $q=m/2+1-n$, where $n=0,1,\cdots$

The general expression of $I(q,r,\sigma)$ is quite complicated; 
a more practical approach is to use a low-$\sigma$ binomial expansion of the kind
\begin{equation}
\sum_{n=1}^{\infty}\frac{n^r}{(n^2+\sigma)^q}=\sum_{l=0}^{\infty}\frac{\sigma^l}{l!}
\frac{\Gamma(-q+1)}{\Gamma(-q-l+1)}\zeta_R (2q+2l-r),
\end{equation}
where $\zeta_R$ is the usual Riemann zeta-function which can be analytically 
continued in the whole complete plane except for the only singular point as $s=1$.
\\
Thus we obtain
\begin{eqnarray} \label{f2}
\lefteqn{ \zeta(s,x|A)  =  \frac{\Omega ^4}{(4\pi)^2} \frac{1}{\sqrt{2\pi}
\Gamma(s)}(\frac{2}{\Omega^2})^s
\Bigg(\sum_{l=0}^{\infty}\frac{(-1)^l\sigma ^l}{l!}
\Gamma (s+l-\frac{1}{2})\zeta _R(2s+2l-3)+{} }\nonumber\\
& & {}+\sum_{k=1}^{\infty}\frac{1}{2^kk!}
\Big[\frac{1}{2k+1}F(s+k-\frac{1}{2};2k+1;
\sigma)+F(s+k-\frac{1}{2};2k;\sigma)+{}
\nonumber\\
& & {}+\sum_{p=0}^{k-1} {2k \choose 2p+1} 2^{2k-2p-1}\frac{B_{2k-2p}}
{k-p}F(s+k-\frac{1}{2};2p+1;\sigma)
\Big]
\Bigg),
\end{eqnarray}
where we have defined
\begin{eqnarray}
F(s+k-\frac{1}{2};r;\sigma)=\sum_{l=0}^{\infty}\sum_{m=0}^{r} {r \choose m}  
\frac{(-1)^{l+m+r}\sigma ^l}{l!}\Gamma (s+k+l-\frac{1}{2})\zeta _R(2s+2k+2l-m-2).
\end{eqnarray}
The expression (\ref{f2}) involves the divergent terms coming from the poles of 
$\Gamma(z)$ at $z=-k ~(k=0,1,2,...)$
\begin{equation}
\Gamma(-k+\epsilon)=\frac{(-1)^k}{k!}(\frac{1}{\epsilon}-\Psi (k+1)+O(\epsilon))
,
\end{equation}
and the pole of $\zeta_R(z)$ at $z=1$
\begin{equation}
\zeta_R(1+2\epsilon)=\frac{1}{2\epsilon}+\gamma,
\end{equation}
where $\gamma$ is Euler-Mascheroni constant, $\gamma=0.5772157$...

The summation above converge whenever Re $s>2$ defining an analytical function 
which can be extended into a meromorphic function defined on the complex $s$
-plane except for two simple poles on the real axis at $s=1$ and $s=2$
\begin{equation} \label{f3}
\zeta (s,x|A)=\zeta_{regular}(s,x|A)+\frac{1}{(4\pi)^2}(\frac{1}{s-2}-
\frac{1}{2}(\sigma+\frac{1}{3})\frac{\Omega^2}{s-1}),
\end{equation}
in concordance with the general theory, since the local Seeley-DeWitt 
coefficients $a_i$ for G\"odel geometry are given by
$a_0=1$ and $a_1=-\frac{\Omega ^2}{2}(\sigma+\frac{1}{3})$.
\\
Also it can be shown that the local zeta-function satisfies the relation
\begin{equation}
\zeta (0,x|A)=\frac{a_2}{4\pi^2}=\frac{\Omega ^2}{(4\pi)^2}
(\frac{4}{45}+\frac{1}{4}(\sigma+\frac{1}{3})^2).
\end{equation}
In deducing the above expressions we have used the relations
\begin{eqnarray}
\sum_{k=1}^{\infty}\frac{\Gamma(k+\frac{1}{2})}{2^kk!} & = & (\sqrt{2}-1)\sqrt{\pi},
\nonumber\\
\sum_{k=1}^{\infty}\frac{\Gamma(k+\frac{1}{2})}{2^kk!}k & = & \frac{\sqrt{2\pi}}{2},
\\
\sum_{k=1}^{\infty}\frac{\Gamma(k+\frac{1}{2})}{2^kk!}k^2 & = & \frac{5\sqrt{2\pi}}{4}.\nonumber
\end{eqnarray}
The global zeta-function, defined as
\begin{equation}
\zeta (s|A)=\sum_N \lambda_N^{-s}
\end {equation}
satisfies the relation
\begin{equation} \label{local}
\zeta (s|A)=\frac{16\pi^3}{\Omega ^3}\zeta (s,x|A).
\end {equation}

%%%%%%%%%%%%%%%%%%%%%%%%%%%%%%%%%%%%%%%%%%%%%%%%%%%%%%%%%%%%%%%%%%%%%%%%%%%%%%
%%%%%%                SOME PHYSICAL APPLICATIONS
%%%%%%%%%%%%%%%%%%%%%%%%%%%%%%%%%%%%%%%%%%%%%%%%%%%%%%%%%%%%%%%%%%%%%%%%%%%%%%
\section{Some physical applications}

Two very useful quantities in the study of quantum effects in curved 
spacetimes are $<\Phi ^2>$ and $S_{eff}$, where $S_{eff}$ is one loop effective action of 
the scalar field $\Phi$ . $<\Phi ^2>$ is a useful quantum quantity because it 
gives one qualitative information about the renormalized stress tensor $<T_{k}^i>$ 
%(where $<T_{ik}>$ 
%is the stress energy tensor operator) 
and can often be computed with much less effort. 
$<\Phi ^2>$ also provides information about spontaneous symmetry 
breaking in a given background.

In the path integral approach, the effective action for a scalar field 
can be formally expressed as the functional determinant of the operator $A$ as
\begin{equation}
S_{eff}[\Phi,g]=-\frac{1}{2}\ln \det(\frac{A}{\mu ^2}),
\end{equation}
where $\mu$ is a scale parameter necessary from dimensional considerations 
(this parameter may remain in the final results and thus can be reabsorbed 
into the renormalized gravitational constant as well as other physically 
measurable parameters).

This determinant is however a formally divergent quantity and needs to be regularized. 
In the framework of the $\zeta$-function renormalization, 
the regularized determinant reads \cite{Moretti:1997qn}
\begin{equation}
S_{eff}[\Phi,g]=-\frac{1}{2}\frac{d}{ds}\bigg| _{s=0} \zeta (s|\frac{A}{\mu ^2})
=-\frac{1}{2}\frac{d}{ds} \bigg| _{s=0} \zeta (s|A)-\frac{1}{2}\frac{d}{ds}
\bigg| _{s=0} \zeta (0|A)\ln \mu ^2.
\end{equation}
For our application, we are interested in $\zeta '(0|A)$. 
To compute the derivative of the global zeta-function in $s=0$, 
we note that if $f(s)$ is a smooth function of $s$, then
\begin{equation}
\frac{d}{ds}\left(\frac{f(s)}{\Gamma (s)} \right)\bigg| _{s=0}=f(0).
\end{equation}
By using the relations (\ref{f2}),(\ref{local}) and the expansion
\begin{displaymath}
\frac{1}{\Gamma(s)} \sim s+\gamma s^2+O(s^3) 
\end{displaymath}
as $s\to 0$, with a bit of algebra, one obtains
\begin{equation}
S_{eff}=-\frac{1}{2}\left( T+\frac{3\gamma}{2}\frac{a_2}{(4\pi)^2}+
\frac{a_2}{(4\pi)^2}\ln\frac{2}{\mu^2\Omega^2}+\frac{1}{2}
\frac{\Omega^4}{(4\pi)^2}\frac{1}{\sqrt{2\pi}}(c_1+c_2+c_3\sigma ^2) \right),
\end{equation}
where
\begin{eqnarray}
T  &=&  \frac{\Omega ^4}{(4\pi)^2} \frac{1}{\sqrt{2\pi}}
(-\frac{\sqrt{\pi}}{240}+\frac{\sqrt{\pi}}{12}\sigma 
+\sum_{l=3}^{\infty}\frac{(-1)^l\sigma ^l}{l!}\Gamma (l-\frac{1}{2})\zeta _R(2l-3)+
\nonumber\\
&+&\sum_{k=1}^{\infty}\frac{1}{2^k k!} \bigg[ \frac{1}{2k+1}
\sum_{l=0}^{\infty}
%\sum_{\scriptstyle l=0\atop\scriptstyle \{l,m\}\ne \{M_1\}}^\infty
% \sum_{m=0}^{2k+1}
 \sum_{\scriptstyle m=0\atop\scriptstyle (l,m)\ne \{M_1\}}^{2k+1}
{2k+1 \choose m}
\frac{(-1)^{l+m+1}\sigma ^l}{l!}\Gamma (k+l-\frac{1}{2})
\zeta _R(2k+2l-m-2)
\nonumber\\
&+& 
\sum_{l=0}^{\infty} 
%\sum_{m=0}^{2k}
 \sum_{\scriptstyle m=0\atop\scriptstyle (l,m)\ne \{M_2\}}^{2k}
{2k \choose m} \frac{(-1)^{l+m}\sigma ^l}{l!}\Gamma (k+l-\frac{1}{2})
\zeta _R(2k+2l-m-2)
\nonumber\\
&+&\sum_{l=0}^{\infty}
%\sum_{p=0}^{k-1}
 \sum_{\scriptstyle p=0\atop\scriptstyle (l,m,p)\ne \{M_3\}}^{k-1}
\sum_{m=0}^{2p+1}
{2k \choose 2p+1} 
\frac{(-1)^{l+m+1} \sigma ^l}{l!} \frac{ B_{2k-2p}}{k-p}
\Gamma (k+l-\frac{1}{2})\zeta _R(2k+2l-m-2)
\bigg],
\end{eqnarray}
and $c_1=0.310856;~c_2=1.17983;~c_3=0.1721$.
Here the indices corresponding to divergent terms are omitted and therefore 
$M_1=\{(0,2k-3),(1,2k-1),(2,2k+1)\},~
M_2=\{(0,2k-3),(1,2k-1)\},
M_3=\{(0,k-2,2p+1),(0,k-1,2p-1),(1,k-1,2p+1)\}$.

The scale $\mu$ represents the usual ambiguity due to a remaining 
finite renormalization and it remains into the final result whenever
another fixed scale is already present in the theory.
The Newton constant $G$ and $1/\Omega$ determine two different scales in the theory.
The constant $1/\Omega$ (related to the scalar curvature via $R=2 \Omega^2$)
is a large distance (cosmological) scale while the distance $l_{Pl}\sim G^{1/2}$
is the Planck scale of the theory.
As expected, the parameter $\Omega$ combines with the renormalization scale $\mu$
to leave an dimensionless argument for the logarithm.

The vacuum expectation value of the field fluctuations 
can be computed within the $\zeta$-function 
regularization scheme by means of the formula
\begin{equation}
<\Phi^2(x)>=\frac{d}{ds}\bigg| _{s=0}
s \zeta (s+1,x|A)+s\zeta(s+1,x|A)\mid_{s=0}\ln \mu^2,
\end{equation}
where $\zeta(s,x|a)$ is the local zeta-function and 
$\mu$ is again the renormalization mass scale. 
Recently it has been shown that this procedure leads to 
the same results as the point-splitting technique \cite{Moretti:1999rf}.

From the expression (\ref{f1}),(\ref{f3}) the behavior of the local zeta-function 
and its derivative near $s=1$ is a simple matter of algebra, 
and we can easily obtain the expectation value of a field fluctuations
\begin{eqnarray}
<\Phi ^2> &=& \frac{\Omega ^2}{(4\pi)^2} \frac{1}{\sqrt{2\pi}}
(-\frac{\sqrt{\pi}}{12}
+\sum_{l=2}^{\infty}\frac{(-1)^l\sigma ^l}{l!}\Gamma (l+\frac{1}{2})\zeta _R(2l-1)+
\nonumber\\
&+&\sum_{k=1}^{\infty}\frac{1}{2^k k!} \bigg[ \frac{1}{2k+1}
\sum_{l=0}^{\infty}
%\sum_{\scriptstyle l=0\atop\scriptstyle \{l,m\}\ne \{M_1\}}^\infty
 \sum_{\scriptstyle m=0\atop\scriptstyle (l,m)\ne \{N_1\}}^{2k+1}
{2k+1 \choose m}
\frac{(-1)^{l+m+1}\sigma ^l}{l!}\Gamma (k+l+\frac{1}{2})
\zeta _R(2k+2l-m)
\nonumber\\
&+& 
\sum_{l=0}^{\infty} 
 \sum_{\scriptstyle m=0\atop\scriptstyle (l,m)\ne (0,2k-1)}^{2k}
{2k \choose m} \frac{(-1)^{l+m}\sigma ^l}{l!}\Gamma (k+l+\frac{1}{2})
\zeta _R(2k+2l-m)
\nonumber\\
&+&\sum_{l=0}^{\infty}
  \sum_{\scriptstyle p=0\atop\scriptstyle (l,p,m)\ne (0,k-1,2p+1) }^{k-1}
\sum_{m=0}^{2p+1}
{2k \choose 2p+1}
2^{2k-2p-1} 
\frac{ B_{2k-2p}(-1)^{l+m+1} \sigma ^l}{(k-p)l!} 
\Gamma (k+l+\frac{1}{2})\zeta _R(2k+2l-m)
\bigg]
\nonumber\\
&+&\frac{3a_1\gamma}{(4\pi)^2}+\frac{a_1}{(4\pi)^2}\log\frac{2\mu^2}{\Omega^2}
+\frac{1}{(4\pi)^2}(\alpha_0+\alpha_1 \sigma),
\end{eqnarray}
where $\alpha_0=-0.188269,~\alpha_1=-0.152411$ and $N_1=\{(0,2k-1),(1,2k+1)\}$. 
Again, the parameter $\Omega$ and the renormalization scale 
$\mu$ combine to leave an dimensionless argument for the logarithm.

In a ``super-$\zeta$-regular theory'' $(\sigma=-1/3)$ 
the local zeta-functional and its derivative are finite in $s=1$; 
the field fluctuations are simply given by the value in $s=1$ of 
the local zeta-function and the above formula reduces to the more 
usual expression also expected from Green function analysis
\begin{equation}
<\Phi^2>=\zeta(1,x|A).
\end{equation}
Both $<\Phi^2>$ and $S_{eff}$ are well behaved quantities and we can
find their values for a given $\sigma$ by performing the above summation.
Although impressive in appearance, these series are quickly convergent
in practice.

Now it is quite straightforward to obtain the partition function and 
then all other thermodynamically quantities for a scalar field. 
The free energy is related to the canonical partition function 
by means of equation $F=-\frac{1}{\beta}\ln Z$. 
The entropy and the internal energy of the system are given 
by the usual thermodynamical 
formula $U=<E>=\frac{\partial}{\partial \beta}(\beta F)$ 
and $S=\beta^2\frac{\partial}{\partial \beta}F$. 
In this way it is possible to obtain exact results beyond the 
WKB approximation discussed in \cite{Radu:1997ds}. 
After all calculation with the Euclideanized G\"odel 
universe have been completed, 
we analytically continue the results back to real values 
of $\Omega$, obtaining results valid on the Lorentzian section.
%\end{section}

%%%%%%%%%%%%%%%%%%%%%%%%%%%%%%%%%%%%%%%%%%%%%%%%%%%%%%%%%%%%%%%%%%%%%%%%%%%%%%
%%%%%%                  FINAL REMARKS
%%%%%%%%%%%%%%%%%%%%%%%%%%%%%%%%%%%%%%%%%%%%%%%%%%%%%%%%%%%%%%%%%%%%%%%%%%%%%%
\section{Final remarks. Conclusions}
In this paper we have obtained exact expressions at one loop for the 
effective action and the vacuum expectation values 
for a scalar field propagating in G\"odel spacetime. Our expression hold for 
massless as well as massive fields, with an arbitrary coupling  with the scalar curvature. 
Similar analysis can be carried out for higher spin fields.

The computation presented here  is only the first step to the study of physically more 
interesting effects. Once $<\Phi^2>$ is known it is easy to compute $<T>$ 
(the expectation value of the trace of the stress-energy tensor) 
for a conformally coupled massive scalar field.  
Since $T=-M^2<\Phi^2>$ for this field and $<\Phi^2>$ is a constant, 
the renormalized value of $<T>$ is just the trace anomaly plus $<\Phi^2>$, $i.e.$
\begin{equation}
<T>=\frac{a_2}{(4\pi)^2}-M^2<\Phi^2>.
\end{equation}
Given the high symmetry of the manifold, it is very convenient to compute 
the renormalized stress energy tensor by using the same formalism of 
``the direct local zeta-function approach'' within the one loop renormalized 
stress-energy tensor of a scalar field is \cite{Moretti:1997qn}
\begin{eqnarray}
\lefteqn{ <T_{ab}(x)>=\{\zeta_{ab}(s+1,x|A)+\frac{1}{2}g_{ab}\zeta (s,x|A)+{} } \\
& & {}+s
\Big[\zeta '_{ab}(s+1,x|A+\ln(\mu^2)\zeta_{ab}(s+1,x|A)
\Big]\}| _{s=0}\nonumber.
\end{eqnarray}
The tensorial zeta-function 
$\zeta(s,x|A)_{ab}$
 is obtained through the 
$s$-analytic continuation in the whole complex plane of the series
\begin{equation}
\zeta_{ab}(s,x|A)=\overline\zeta_{ab}(s,x|A)+\xi R_{ab}(x)
\zeta(s,x|A)-\frac{1}{2}\zeta(s-1,x|A)
\end{equation} 
with
\begin{equation}
\overline\zeta_{ab}(s,x|A)=\sum_N \lambda_N^{-s}\nabla _a\Phi _N^*\nabla_b\Phi _N(x),
\end{equation} 
where the homogeneity of the background was taken into account.
By using the properties of Wigner functions, it is possible to find the renormalized 
$\zeta(s,x|A)_{ab}$. 
Since the calculations are very difficult, the explicit 
computation of $<T_k^i>$
will be considered elsewhere.
The $zz$ component of the renormalized stress tensor is the only one that 
can be easily computed by taking into account the relation
\begin{equation}
\overline\zeta_{zz}(s,x|A)=\frac{\Omega^2}{2(s-1)}\zeta (s-1,x|A).
\end{equation} 
Hence
\begin{equation}
<T_{zz}(x)>=\frac{1}{2}\left(\frac{d}{ds}\bigg| _{s=0}\zeta(s,x|A)+\zeta(0,x|A)\ln(\mu^2)
\right).
\end{equation}
However, few remarks can be given even at this stage. 
In general, it seems that divergences 
in the energy momentum-tensor occur when one has closed or self-interacting 
null geodesics \cite{Hawking:1992nk}. 
However, in G\"odel universe there is no Cauchy horizon 
containing closed null geodesics. Also from the analysis of free motion in 
G\"odel spacetime, we know that the
geodesic motion do not follow a CTC \cite{Chandrasekhar,Novello}.

We find that both the vacuum fluctuations of the field 
and the renormalized stress energy-tensor 
$<T_k^i>$ are well behaved quantities in every spacetime point. 
Thus CTCs do not mean that the energy-momentum tensor must diverge and 
we cannot hope the causality violation which occurs in G\"odel spacetime 
might be removed by quantum effects. 
A complete answer is not possible until the investigation of the problem 
of back reaction on the metric.
\\
\newline
{\bf Acknowledgement}
\newline
The help of prof. E. Elizalde in a early stage of 
this work is gratefully acknowledged.
\newline
The work of one of the authors (E.R.) was performed in the context of the
Graduiertenkolleg of the Deutsche Forschungsgemeinschaft (DFG):
Nichtlineare Differentialgleichungen: Modellierung,Theorie, Numerik, Visualisierung.

%%%%%%%%%%%%%%%%%%%%%%%%%%%%%%%%%%%%%%%%%%%%%%%%%%%%%%%%%%%%%%%%%%%%%%%%%%%%%%
%%%%%%                  APPENDIX
%%%%%%%%%%%%%%%%%%%%%%%%%%%%%%%%%%%%%%%%%%%%%%%%%%%%%%%%%%%%%%%%%%%%%%%%%%%%%%
\section{Appendix. Definition and evaluation of the $I(q,p,\sigma)$ function}
In this section we shall study the function $I(q,r,\sigma)$
\begin{equation} \label{a1}
I(q,r,\sigma)=\sum_{n=1}^{\infty}\frac{n(n-1)^r}{(n^2+\sigma)^q}.
\end{equation}
This series is a generalization of the one examined by Shen et al. \cite{Shen:1985ir}. 
Following the same procedure, we find that $I(q,r,\sigma)$ 
have poles at $r=\frac{m}{2}-n+1$. Employing the Plana sum 
formula $I(q,r,\sigma)$ can be written as
\begin{eqnarray} \label{a2}
\lefteqn { I(q,r,\sigma)=\sum_{m=0}^{r}{r \choose m}
(-1)^{r+m}\int_{0}^{\infty}dx\frac{x^{m+1}}{(x^2+\sigma)^q}
-\int_{0}^{1}dx\frac{x(x-1)^{r}}{(x^2+\sigma)^q}+{} } 
\\
& & {}+i^{r+1}\int_{0}^{\infty}dt\frac{t^r}{e^{2\pi t}-1}
\Big(
\frac{1+it}{((1+it)^2+\sigma)^q}+\frac{(-1)^{r+1}(1-it)}{((1+it)^2+\sigma)^q}
\Big)\nonumber.
\end{eqnarray}
The integral
\begin{equation} \label{a3}
\int_{0}^{\infty}dx\frac{x^{m+1}}{(x^2+\sigma)^q}=\frac{\sigma
^{\frac{m}{2}+1-q}}{2}\frac{\Gamma(\frac{m}{2}+1)\Gamma(q-\frac{m}{2}-1)}{\Gamma(q)}
\end{equation}
vanishes at negative integer $q$, but diverges at $q=\frac{m}{2}+1-n$, 
where $n=0,1,2,...$ 

This is the only divergent part on the right hand side of eq. (\ref{a2}). 
One can expand the singular term in a series about the pole of the Gamma 
function at $q=\frac{m}{2}+1-n$ 
\begin{equation}
\int_{0}^{\infty}dx\frac{x^{m+1}}{(x^2+\sigma)^q}
=\frac{\sigma^n}{2}
\frac{\Gamma(\frac{m}{2}+1)}{\Gamma(\frac{m}{2}+1-n)}
\Bigg[\frac{(-1)^n}
{n!}\frac{1}{\nu}+\Psi (n+1)
+\frac{1}{2}\nu
\Big(\frac{\pi^2}{3}+\Psi^2(n+1)-\Psi'(n+1)
\Big)
\Bigg]
\end{equation}
where $\Psi(x)=\frac{d}{dx}\ln\Gamma(x)$. 
We denote the singular part of the function 
$I(q,r,\sigma)$ by $I_{-1}$ and the regular part by $I_0$.
Therefore
\begin{eqnarray}
I_{-1}(n,r,\sigma)&=&\sum_{m=0}^{r}{r \choose m}(-1)^{r+m}
\frac{\sigma^n}{2}\frac{\Gamma(\frac{m}{2}+1)}{\Gamma(\frac{m}{2}+1-n)}
\frac{(-1)^n}{n!},
\nonumber\\
I_0(n,r,\sigma)&=&
\sum_{m=0}^{r}\Big[{r \choose m}(-1)^{r+m}\frac{\sigma^n}{2}
\frac{\Gamma(\frac{m}{2}+1)}{\Gamma(\frac{m}{2}+1-n)}\Psi(n+1)
-\int_{0}^{1}dx\frac{x(x-1)^r}{(x^2+\sigma)^{\frac{m}{2}+1-n}}
%+{} }\nonumber 
\nonumber\\
&+&i^{r+1}\int_{0}^{\infty}dt
\frac{t^r}{e^{2\pi t}-1}
\left (\frac{1+it}{((1+it)^2+\sigma)^{\frac{m}{2}+1-n}}
+\frac{(-1)^{r+1}(1-it)}
      {((1-it)^2+\sigma)^{\frac{m}{2}+1-n}
                       } \right)\Big].
\end{eqnarray}
An analytic result for $I_0(n,r,\sigma),~ I_{-1}(n,r,\sigma)$ is not accessible; 
however we may seek the asymptotic expansions for particular ranges 
of $\sigma$. Note that for $\sigma=1$, the $n=1$ term 
is excluded from the above summation.

%%%%%%%%%%%%%%%%%%%%%%%%%%%%%%%%%%%%%%%%%%%%%%%%%%%%%%%%%%%%%%%%%%%%%%%%%%%%%%


\begin{thebibliography}{99}
%%%%%%%%%%%%%%%%%%%%%%%%%%%%%%%%%%%%%%%%%%%%%%%%%%%%%%%%%%%%%%%%%%%%%%%%%%%%%%
%\cite{Godel:1949ga}
\bibitem{Godel:1949ga}
K.~G\"odel,
%``An Example Of A New Type Of Cosmological Solutions Of Einstein's 
%Field Equations Of Graviation,''
\emph{Rev.\ Mod.\ Phys.}\  {\bf 21}, 447 (1949).
%%%%%%%%%%%%%%%%%%%%%%%%%%%%%%%%%%%%%%%%%%%%%%%%%%%%%%%%%%%%%%%%%%%%%%%%%%%%%
%\cite{reboucas}
\bibitem{reboucas}
M.~J. Rebou\c{c}as and J. Tiomno,
\emph{Phys.\ Rev.}\ D {\bf 28}, 1251 (1983).
%%CITATION = PHRVA,D28,1251;%%
%%%%%%%%%%%%%%%%%%%%%%%%%%%%%%%%%%%%%%%%%%%%%%%%%%%%%%%%%%%%%%%%%%%%%%%%%%%%%%
%\cite{Som}
\bibitem{Som}
M.~M.~Som and A.~K.~Raychaudhuri,  
\emph{Proc. R. Soc. London} A {\bf304}, 81 (1968). 
%%%%%%%%%%%%%%%%%%%%%%%%%%%%%%%%%%%%%%%%%%%%%%%%%%%%%%%%%%%%%%%%%%%%%%%%%%%%%%
%\cite{Sasse}
%\bibitem{Sasse}
%F.~D. Sasse, I.~D.~Soares and J.~ Tiomno,
%Braz. J. Phys.  {\bf 25}, 204 (1995).
%%%%%%%%%%%%%%%%%%%%%%%%%%%%%%%%%%%%%%%%%%%%%%%%%%%%%%%%%%%%%%%%%%%%%%%%%%%%%%
%\cite{Chandrasekhar}
\bibitem{Chandrasekhar}
S.~Chandrasekhar and J.~P.~Wright
\emph{Proc. Nat. Acad. Sci.}  {\bf 47}, 34 (1961).
%%%%%%%%%%%%%%%%%%%%%%%%%%%%%%%%%%%%%%%%%%%%%%%%%%%%%%%%%%%%%%%%%%%%%%%%%%%%%%
%\cite{Novello}
\bibitem{Novello}
M.~Novello, I.~D.~Soares and J.~ Tiomno,
\emph{Phys.\ Rev.\ }D {\bf 27}, 779 (1983).
%%%%%%%%%%%%%%%%%%%%%%%%%%%%%%%%%%%%%%%%%%%%%%%%%%%%%%%%%%%%%%%%%%%%%%%%%%%%%%
\bibitem{Hawking}
S.~W.~ Hawking,~ G. ~F.~ R.~ Ellis,
\emph{The large structure of space-time}
(Cambridge, Cambridge University Press, 1973), Section 5.7.
%%%%%%%%%%%%%%%%%%%%%%%%%%%%%%%%%%%%%%%%%%%%%%%%%%%%%%%%%%%%%%%%%%%%%%%%%%%%%%
%\cite{Chimento:1989ax}
\bibitem{Chimento:1989ax}
L.~P.~Chimento, A.~S.~Jakubi and J.~Pullin,
%``Coleman-Weinberg Symmetry Breaking In A Rotating Space-Time,''
\emph{Class.\ Quant.\ Grav.}\  {\bf 6}, L45 (1989).
%%CITATION = CQGRD,6,L45;%%
%%%%%%%%%%%%%%%%%%%%%%%%%%%%%%%%%%%%%%%%%%%%%%%%%%%%%%%%%%%%%%%%%%%%%%%%%%%%%%
%\cite{Huang:1991eg}
\bibitem{Huang:1991eg}
W.~H.~Huang,
%``Finite temperature cosmological phase transition in a rotating space-time,''
\emph{Class.\ Quant.\ Grav.} {\bf 8}, 1471 (1991).
%%CITATION = CQGRD,8,1471;%%
%%%%%%%%%%%%%%%%%%%%%%%%%%%%%%%%%%%%%%%%%%%%%%%%%%%%%%%%%%%%%%%%%%%%%%%%%%%%%%
%\cite{Radu:1998sk}
\bibitem{Radu:1998sk}
E.~Radu,
%``On the Euclidean approach to quantum field theory in Goedel space-time,''
\emph{Phys. Lett.}A {\bf 247}, 207 (1998).
%%CITATION = PHLTA,A247,207;%%
%%%%%%%%%%%%%%%%%%%%%%%%%%%%%%%%%%%%%%%%%%%%%%%%%%%%%%%%%%%%%%%%%%%%%%%%%%%%%%
%\cite{Leahy:1982dj}
\bibitem{Leahy:1982dj}
D.~A.~Leahy,
%``Scalar And Neutrino Fields In The Godel Universe. (Talk),''
\emph{Int. J. Theor. Phys.}  {\bf 21}, 703 (1982).
%%CITATION = IJTPB,21,703;%%
%%%%%%%%%%%%%%%%%%%%%%%%%%%%%%%%%%%%%%%%%%%%%%%%%%%%%%%%%%%%%%%%%%%%%%%%%%%%%%
%\cite{Novello1993}
\bibitem{Novello1993}
M.~Novello, N.~F.~Svaiter and M.~E.~X.~Guimaraes,
\emph{Gen. Rel. Grav.} {\bf 25}, 137 (1993).
%%CITATION = PHRVA,D49,825;%%
%%%%%%%%%%%%%%%%%%%%%%%%%%%%%%%%%%%%%%%%%%%%%%%%%%%%%%%%%%%%%%%%%%%%%%%%%%%%%%
%\cite{Carneiro:2000yw}
\bibitem{Carneiro:2000yw}
S.~Carneiro,
%``A Goedel-Friedman cosmology?,''
\emph{Phys. Rev.}D {\bf 61}, 083506 (2000).
%[gr-qc/9809043].
%%CITATION = GR-QC 9809043;%%
%%%%%%%%%%%%%%%%%%%%%%%%%%%%%%%%%%%%%%%%%%%%%%%%%%%%%%%%%%%%%%%%%%%%%%%%%%%%%%
%\cite{Obukhov:2000jf}
\bibitem{Obukhov:2000jf}
Y.~N.~Obukhov,
%``On physical foundations and observational effects of cosmic rotation,''
astro-ph/0008106.
%%CITATION = ASTRO-PH 0008106;%%
%%%%%%%%%%%%%%%%%%%%%%%%%%%%%%%%%%%%%%%%%%%%%%%%%%%%%%%%%%%%%%%%%%%%%%%%%%%%%%
%\cite{Chamblin:1999pz}
\bibitem{Chamblin:1999pz}
A.~Chamblin, R.~Emparan, C.~V.~Johnson and R.~C.~Myers,
%``Large N phases, gravitational instantons and the nuts and bolts of AdS  holography,''
\emph{Phys. Rev.} D {\bf 59}, 064010 (1999).
%[hep-th/9808177].
%%CITATION = HEP-TH 9808177;%%
%%%%%%%%%%%%%%%%%%%%%%%%%%%%%%%%%%%%%%%%%%%%%%%%%%%%%%%%%%%%%%%%%%%%%%%%%%%%%%
%\cite{Maldacena:1998re}
\bibitem{Maldacena:1998re}
J.~Maldacena,
%``The large N limit of superconformal field theories and supergravity,''
\emph{Adv. Theor. Math. Phys.}  {\bf 2}, 231 (1998).
%[Int.\ J.\ Theor.\ Phys.\  {\bf 38} (1998) 1113]
%[hep-th/9711200].
%%CITATION = HEP-TH 9711200;%%
%%%%%%%%%%%%%%%%%%%%%%%%%%%%%%%%%%%%%%%%%%%%%%%%%%%%%%%%%%%%%%%%%%%%%%%%%%%%%%
%\cite{Hawking:1995zi}
\bibitem{Hawking:1995zi}
S.~W.~Hawking,
%``Quantum coherence and closed timelike curves,''
\emph{Phys. Rev.} D {\bf 52}, 5681 (1995).
%[gr-qc/9502017].
%%CITATION = GR-QC 9502017;%%
%%%%%%%%%%%%%%%%%%%%%%%%%%%%%%%%%%%%%%%%%%%%%%%%%%%%%%%%%%%%%%%%%%%%%%%%%%%%%%
%\cite{Gibbons:1977ue}
\bibitem{Gibbons:1977ue}
G.~W.~Gibbons and S.~W.~Hawking,
%``Action Integrals And Partition Functions In Quantum Gravity,''
\emph{Phys. Rev.} D {\bf 15}, 2752 (1977).
%%CITATION = PHRVA,D15,2752;%%
%%%%%%%%%%%%%%%%%%%%%%%%%%%%%%%%%%%%%%%%%%%%%%%%%%%%%%%%%%%%%%%%%%%%%%%%%%%%%%
%\cite{Rooman:1998xf}
\bibitem{Rooman:1998xf}
M.~Rooman and P.~Spindel,
%``Goedel metric as a squashed anti-de Sitter geometry,''
\emph{Class. Quant. Grav.}  {\bf 15}, 3241 (1998).
%[gr-qc/9804027].
%%CITATION = GR-QC 9804027;%
%%%%%%%%%%%%%%%%%%%%%%%%%%%%%%%%%%%%%%%%%%%%%%%%%%%%%%%%%%%%%%%%%%%%%%%%%%%%%%
%\cite{Taub:1951ez}
\bibitem{Taub:1951ez}
A.~H.~Taub,
%``Empty Space-Times Admitting A Three Parameter Group Of Motions,''
\emph{Annals Math.}  {\bf 53}, 472 (1951).
%%CITATION = ANMAA,53,472;%%
%%%%%%%%%%%%%%%%%%%%%%%%%%%%%%%%%%%%%%%%%%%%%%%%%%%%%%%%%%%%%%%%%%%%%%%%%%%%%%
%\cite{Iellici:1998ce}
\bibitem{Iellici:1998ce}
D.~Iellici,
%``Aspects and applications of quantum field theory on spaces with conical  singularities,''
gr-qc/9805058.
%%CITATION = GR-QC 9805058;%%
%%%%%%%%%%%%%%%%%%%%%%%%%%%%%%%%%%%%%%%%%%%%%%%%%%%%%%%%%%%%%%%%%%%%%%%%%%%%%%
%\cite{Critchley:1981am}
\bibitem{Critchley:1981am}
R.~Critchley and J.~S.~Dowker,
%``Vacuum Stress Tensor For A Slightly Squashed Einstein Universe,''
\emph{J. Phys.}A {\bf 14}, 1943 (1981).
%%CITATION = JPAGB,A14,1943;%%
%%%%%%%%%%%%%%%%%%%%%%%%%%%%%%%%%%%%%%%%%%%%%%%%%%%%%%%%%%%%%%%%%%%%%%%%%%%%%%
%\cite{Shen:1985ir}
\bibitem{Shen:1985ir}
T.~C.~Shen, B.~L.~Hu and D.~J.~O'Connor,
%``Symmetry Behavior Of The Static Taub Universe: Effect Of Curvature Anisotropy,''
\emph{Phys. Rev.}D {\bf 31}, 2401 (1985).
%%CITATION = PHRVA,D31,2401;%%
%%%%%%%%%%%%%%%%%%%%%%%%%%%%%%%%%%%%%%%%%%%%%%%%%%%%%%%%%%%%%%%%%%%%%%%%%%%%%%
%\cite{Stylianopoulos:1989sk}
\bibitem{Stylianopoulos:1989sk}
A.~Stylianopoulos,
%``Finite Temperature Symmetry Breaking In An Anisotropic Universe,''
\emph{Phys. Rev.} D {\bf 40}, 3319 (1989).
%%CITATION = PHRVA,D40,3319;%%
%%%%%%%%%%%%%%%%%%%%%%%%%%%%%%%%%%%%%%%%%%%%%%%%%%%%%%%%%%%%%%%%%%%%%%%%%%%%%%
%\cite{Dowker:1999pi}
\bibitem{Dowker:1999pi}
J.~S.~Dowker,
%``Effective actions on the squashed three-sphere,''
\emph{Class. Quant. Grav.}  {\bf 16}, 1937 (1999).
%[hep-th/9812202].
%%CITATION = HEP-TH 9812202;%%
%%%%%%%%%%%%%%%%%%%%%%%%%%%%%%%%%%%%%%%%%%%%%%%%%%%%%%%%%%%%%%%%%%%%%%%%%%%%%%
%\cite{DeFrancia:2001xm}
\bibitem{DeFrancia:2001xm}
M.~De Francia, K.~Kirsten and J.~S.~Dowker,
%``Effective actions on squashed lens spaces,''
\emph{Class. Quant. Grav.}  {\bf 18}, 955 (2001).
%[hep-th/0008059].
%%CITATION = HEP-TH 0008059;%%
%%%%%%%%%%%%%%%%%%%%%%%%%%%%%%%%%%%%%%%%%%%%%%%%%%%%%%%%%%%%%%%%%%%%%%%%%%%%%%
%\cite{Moretti:1997qn}
\bibitem{Moretti:1997qn}
V.~Moretti,
%``Direct zeta-function approach and renormalization of one-loop stress  
%tensors in curved spacetimes,''
\emph{Phys. Rev.} D {\bf 56}, 7797 (1997).
%[hep-th/9705060].
%%CITATION = HEP-TH 9705060;%%
%%%%%%%%%%%%%%%%%%%%%%%%%%%%%%%%%%%%%%%%%%%%%%%%%%%%%%%%%%%%%%%%%%%%%%%%%%%%%%
%\cite{Ghika:1978vq}
%\bibitem{Ghika:1978vq}
%G.~Ghika and M.~Visinescu,
%``Zeta Function Regularization Of The One Loop Effective Potential,''
%\emph{Nuovo Cim.} A {\bf 46}, 25 (1978).
%%CITATION = NUCIA,A46,25;%%
%%%%%%%%%%%%%%%%%%%%%%%%%%%%%%%%%%%%%%%%%%%%%%%%%%%%%%%%%%%%%%%%%%%%%%%%%%%%%%
%\cite{Moretti:1999rf}
\bibitem{Moretti:1999rf}
V.~Moretti,
%``Local zeta-function techniques vs point-splitting procedure: A few  rigorous results,''
\emph{Commun. Math. Phys.}\  {\bf 201}, 327 (1999).
%[gr-qc/9805091].
%%CITATION = GR-QC 9805091;%%
%%%%%%%%%%%%%%%%%%%%%%%%%%%%%%%%%%%%%%%%%%%%%%%%%%%%%%%%%%%%%%%%%%%%%%%%%%%%%%
%%\cite{Radu:1997ds}
\bibitem{Radu:1997ds}
E.~Radu,
%``A note on the thermodynamic properties of a scalar field in a rotating  space-time,''
\emph{Mod. Phys. Lett.}\ A {\bf 12}, 2341 (1997).
%%CITATION = MPLAE,A12,2341;
%%%%%%%%%%%%%%%%%%%%%%%%%%%%%%%%%%%%%%%%%%%%%%%%%%%%%%%%%%%%%%%%%%%%%%%%%%%%%%%
%\cite{Hawking:1992nk}
\bibitem{Hawking:1992nk}
S.~W.~Hawking,
%``The Chronology protection conjecture,''
\emph{Phys. Rev.} D {\bf 46}, 603 (1992).
%%CITATION = PHRVA,D46,603;%%
%%%%%%%%%%%%%%%%%%%%%%%%%%%%%%%%%%%%%%%%%%%%%%%%%%%%%%%%%%%%%%%%%%%%%%%%%%%%%%
\end{thebibliography}
\end{document}